# A thermodynamic approach of the mechano-chemical coupling during the oxidation of uranium dioxide

Nicolas Creton[1, a], Virgil Optasanu[1, b], Tony Montesin[1, c], Sébastien Garruchet[1, d] and Lionel Desgranges[2, e]

[1]I.C.B., UMR 5209 CNRS, 9 Av. Alain Savary, BP 47870, 21078 Dijon France,

[2]CEA / DEN/DEC/SESC/LLCC, CEA Cadarache, 13115 St Paul Lez Durance, France

[a]nicolas.creton@u-bourgogne.fr, [b]virgil.optasanu@u-bourgogne.fr, [c]tony.montesin@u-bourgogne.fr, [d]sebastien.garruchet@u-bourgogne.fr, [e]lionel.desgranges@cea.fr

**Abstract**. The aim of the present work is to introduce a thermodynamic model to describe the growth of an oxide layer on a metallic substrate. More precisely, this paper offers a study of oxygen dissolution into a solid, and its consequences on the apparition of mechanical stresses. They strongly influence the oxidation processes and may be, in some materials, responsible for cracking. To realize this study, mechanical considerations are introduced into the classical diffusion laws. Simulations were made for the particular case of uranium dioxide, which undergoes the chemical fragmentation. According to our simulations, the hypothesis of a compression stress field into the oxidised $UO_2$ compound near the internal interface is consistent with the interpretation of the mechanisms of oxidation observed experimentally.

**Introduction**

In contact with oxygen, some crystalline solids undergo a chemical transformation during which the cracking and fragmentation of the initial solid is observed. Known as "chemical fragmentation", this oxidation reaction induces mechanical strains due to strong interactions between the different mechanisms that occur during the oxide layer growth. They can be induced by the diffusion of species or by the chemical oxidation reaction. But they influence themselves the diffusion and the reaction, modifying particles flux and interfaces displacement velocities, governed by the thermodynamic forces (chemical potential gradients and affinity), which characterize these processes. In the way of the theoretical approach of such a process, a purely chemical formulation cannot explain by itself a phenomenon like the chemical fragmentation. To solve such a problem, a mechano-chemical approach based on non-equilibrium thermodynamics has been developed [1]. Initially written for the study of Zr anionic oxidation, this model allows us to set a new expression for the equation governing the diffusion of species into the solid, and a new formulation of the substrate/oxide interface displacement including mechanical terms.

In the first part of this paper, thermodynamic bases of our model are reminded. Furthermore, this first part accents on the volume component of the intrinsic dissipation, which leads to the formulation of a new expression for the matter transport law. This one is then applied to an $UO_2$ grain submitted to different solicitation conditions in order to study its mechano-chemical behaviour. Some numerical simulations are realized to validate our equations, by considering some experimental observations of $UO_2$ transformation into $U_4O_9$, $U_3O_7$ or/and $U_3O_8$.

This theoretical approach is the first step of a more global study. It will finally lead to the simulation of the different oxide phases growths at the surface of any material in which the chemical fragmentation occurs.

**Thermodynamics of the diffusion-reaction process**

The construction of an original predictive model concerning an anionic oxidation process needs to obtain the evolution law governing the motion of the substrate/oxide interface, including mechanics and mass transport. For this, non-equilibrium thermodynamics is used to determine the intrinsic dissipation associated both with the diffusion of species into the metal and the motion of the reactive interface.

**Dissipation of the system**. Let's consider a quasi-static and isothermal evolution of an opened thermodynamic system V (Fig. 1) including a moving internal interface $\Sigma$. If $\Phi = E - TS$ is the Helmholtz free energy of the system, the first and second principles leads to the following dissipation equation:

$$\mathcal{D} = P_{ext} - \dot{\Phi} \geq 0 . \tag{1}$$

In this expression, $P_{ext}$ is the power developed by the forces emerging from the mass flux and the external strengths applied to the external boundary of the system. E is the internal energy, T the temperature and S the entropy.

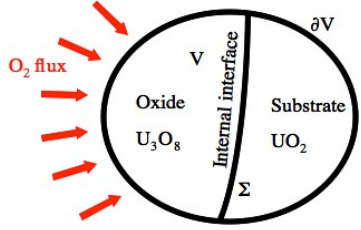

Fig. 1: Schematic representation of the system.

The intrinsic dissipation will allow us to obtain the evolution equations for the internal variables governing the oxidation transformation of the substrate. It is then necessary to find, in Eq.1, an expression for $\Phi$ and $P_{ext}$ introducing the internal variables and constants of the system.

**Helmholtz free energy of the system**. The free energy $\Phi$ is a function of three independent internal variables, strongly coupled:
- the elastic strain $\varepsilon_{ij}^e$,
- the concentration $c$,
- the reaction's degree of conversion $\xi$.

It appears judicious to divide the expression of this energy in two parts, the one in volume to take into account the mechanical properties and diffusion in the system, and the other one in surface to represent the reaction at the interface $\Sigma$:

$$\Phi = \int_V \Phi_V(\varepsilon_{ij}^e, c) dV + \int_\Sigma \Phi_\Sigma(\xi) d\Sigma . \tag{2}$$

By Introducing the Cauchy stress tensor $\sigma_{ij}$, the chemical potential $\mu_\gamma$ of a species $\gamma$ and the chemical affinity A of the reaction as:

$$\sigma_{ij} = \left( \frac{\partial \Phi_V}{\partial \varepsilon_{ij}^e} \right)_c , \mu_\gamma = \left( \frac{\partial \Phi_V}{\partial c} \right)_{\varepsilon_{ij}^e, c_{\beta \neq \gamma}} , A = \left( \frac{\partial \Phi_V}{\partial \xi} \right) . \tag{3}$$

we obtain then a first expression for the Helmholtz free energy [4]:

$$\Phi_V = \int \sigma_{ij} d\varepsilon_{ij}^e + \sum_\gamma \int \mu_\gamma dc_\gamma , \text{ and } \Phi_\Sigma = \int A d\xi . \tag{4}$$

In $\Phi_V$, the first term corresponds to a mechanical contribution representing the differential of elastic energy density. The second term corresponds to a chemical contribution led to the flux of mass diffusing in the system. $\Phi_\Sigma$ represents a chemical contribution due to the reactions existing inside the system.

**Helmholtz free energy evolution.** The evolution of the Helmholtz free energy is due to the diffusion of some species inside the material, whose consequence is to gradually modify the substrate composition. When the limits in concentration of species are reached, the chemical reaction can occur, the oxide grows, and the boundary moves.

To determine the time evolution of the free energy, it is useful to write the time derivative of an integral including discontinuity surfaces $\Sigma$:

$$\dot{\Phi}_V = \int_V \frac{d\Phi_V}{dt} dV + \int_\Sigma [\Phi_V] \vec{\omega}.\vec{n} d\Sigma . \tag{5}$$

where $\vec{\omega}$ represents the speed of propagation of the surface $\Sigma$ and $\vec{n}$ the normal to this surface. The brackets used here correspond to the "jump" of a representative variable $\alpha$ of the system through the internal surface $\Sigma$. For example, if $\alpha^+$ and $\alpha^-$ are the limits of $\alpha$ when the interface $\Sigma$ is approached from the oxide and from the metal respectively, then the jump $[\alpha]$ across the interface is written as: $[\alpha] = \alpha^+ - \alpha^-$. In Eq. 5, $[\Phi_V]$ is the jump of free energy density through the surface $\Sigma$. In the same way, we can write the time derivative of $\Phi_\Sigma$:

$$\dot{\Phi}_\Sigma = \int_\Sigma \frac{d\Phi_\Sigma}{dt} d\Sigma - \int_\Sigma div_\Sigma(\vec{\omega}.\Phi_\Sigma) d\Sigma . \tag{6}$$

These two equations, combined with Eq. 4 leads to an expression of $\dot{\Phi}$ (Eq. 7):

$$\dot{\Phi} = \int_V \left( \sigma_{ij} \dot{\varepsilon}^e_{ij} + \sum_\gamma \mu_\gamma \dot{c}_\gamma \right) dV - \int_\Sigma \left\{ \left( \langle\sigma_{ij}\rangle [\varepsilon^e_{ij}] + \sum_\gamma \int_{c_{\gamma-}}^{c_{\gamma+}} \mu_\gamma dc_\gamma \right) \vec{\omega}.\vec{n} + A\dot{\xi} \right\} d\Sigma . \tag{7}$$

In this expression, $\langle\sigma_{ij}\rangle = \frac{1}{2}\left(\sigma_{ij}^{ox} + \sigma_{ij}^{met}\right)$ represents the average stress and $\left[\varepsilon^e_{ij}\right]$ the strain jump across the interface $\Sigma$.

**Power developed by external forces.** When volume forces are neglected, the oxygen flow through the external surface creates an external force acting on the surface and responsible for the occurence of an external stress ($\sigma_{ij}$) applied to the volume boundary. In other words, the power of external forces (and exclusively surface forces here) is due to:
- the efforts applied to $\partial V$.
- the mass flux entering the system through this surface.

This power can be written as:

$$P_{ext} = \int_{\partial V} v_i.(\sigma_{ij} n_j) dS - \sum_\gamma \int_{\partial V} \mu_\gamma (\vec{J}_\gamma.\vec{n}) dS . \tag{8}$$

In this expression, $v_i$ is a speed field and $J_\gamma$ the mass flux of the constituent $\gamma$ across the surface $\partial V$. Using the divergence operator, it is possible to rewrite this power as:

$$P_{ext} = \int_V \left\{ \sigma_{ij} \dot{\varepsilon}_{ij} - \sum_\gamma \left( \mu_\gamma div(\vec{J}_\gamma) + \vec{J}_\gamma.\vec{\nabla}\mu_\gamma \right) \right\} dV - \int_\Sigma \left\{ \sigma_{ij} [\varepsilon_{ij}] \omega_\alpha n_\alpha + \sum_\gamma [\mu_\gamma \vec{J}_\gamma.\vec{n}] \right\} d\Sigma . \tag{9}$$

**Dissipation calculation**. Let's classically break up the strain into an elastic part and an inelastic part: $\varepsilon_{ij} = \varepsilon_{ij}^e + \varepsilon_{ij}^{inel}$ ($\varepsilon_{ij}^{inel}$ represents all the inelastic strain appearing during the oxidation of metal). Furthermore, the mass conservation relation can be written as:

$$\frac{\partial c_\gamma}{\partial t} = -div\left(\vec{J_\gamma}\right). \tag{10}$$

From Eq. 7 and Eq. 9, we can determine the dissipation expression of the system:

$$\mathcal{D} = \int_V \left\{ \sigma_{ij}\dot{\varepsilon}_{ij}^{inel} - \sum_\gamma \vec{J_\gamma}.\vec{\nabla}\mu_\gamma \right\} dV$$
$$- \int_\Sigma \left\{ \langle\sigma_{ij}\rangle[\varepsilon_{ij}^{inel}]\vec{\omega}.\vec{n} + \sum_\gamma [\mu_\gamma \vec{J_\gamma}].\vec{n} - \left(\sum_\gamma \int_{c_{\gamma-}}^{c_{\gamma+}} \mu_\gamma dc_\gamma\right)\vec{\omega}.\vec{n} + A\dot\xi \right\}d\Sigma. \tag{11}$$

**Volume part of the dissipation.** The first integral in Eq. 11 corresponds to the volume part of the dissipation:

$$\mathcal{D}_V = \int_V \left\{ \sigma_{ij}\dot{\varepsilon}_{ij}^{inel} - \sum_\gamma \vec{J_\gamma}.\vec{\nabla}\mu_\gamma \right\} dV. \tag{12}$$

It represents the energy dissipated in the volume, due to:
- the oxygen diffusion process,
- the plastic strains generated inside the volume by the oxidation process.

This integral should contain an evolution law similar to the Fick's law (in accordance with the hypothesis that the diffusion process takes place in volume). Let's consider the inelastic strain as $\varepsilon_{ij}^{inel} = \varepsilon_{ij}^{ch} + \varepsilon_{ij}^p$, where $\varepsilon_{ij}^{ch}$ corresponds to the deformation of the lattice due to the species diffusion and $\varepsilon_{ij}^p$ corresponds to every other kinds of deformation (plastic part). A definition of the chemical strain was given by Larché [5], who introduced a chemical expansion coefficient $\eta_{ij}^\gamma$ representing the deformation generated by the species γ diffusing in the material, per unit of concentration:

$$\eta_{ij}^\gamma = \left(\frac{\partial \varepsilon_{ij}^{ch}}{\partial c_\gamma}\right)_{\sigma_{ij},T}. \tag{13}$$

In the particular case of UO$_2$, the coefficient $\eta_{ij}^\gamma$ is negative: the volume of UO$_2$ decreases when a species diffuses in it. From Eq. 12 and Eq. 13, and if $\eta_{ij}^\gamma$ does not depend on time and concentration, the volume dissipation becomes:

$$\mathcal{D}_V = \sum_\gamma \int_V \left\{ \sigma_{ij}\dot{\varepsilon}_{ij}^p + \sigma_{ij}\eta_{ij}^\gamma \dot{c}_\gamma - \vec{J_\gamma}.\vec{\nabla}\mu_\gamma \right\} dV. \tag{14}$$

The two last terms of Eq. 14 correspond to the diffusion process inside the material: they must be governed by the same internal variables $\vec{J_\gamma}$. Considering an infinitesimal volume δV and introducing the definition of the flux $\vec{J_\gamma} = c_\gamma \vec{v_\gamma}$ ($\vec{v_\gamma}$ is the velocity of the species γ), we obtain:

$$\mathcal{D}_{\delta V} = \left\{ \sigma_{ij} \dot{\varepsilon}_{ij}^{p} - \sum_{\gamma} \vec{J_{\gamma}} \cdot \left\{ \nabla \left( \mu_{\gamma} + \sigma_{ij} \eta_{ij}^{\gamma} \right) - \left( \frac{\partial \sigma_{ij} \eta_{ij}^{\gamma}}{\partial c_{\gamma}} + \frac{\sigma_{ij} \eta_{ij}^{\gamma}}{c_{\gamma}} \right) \vec{\nabla} c_{\gamma} \right\} \right\}. \tag{15}$$

As the dissipation can be written as the sum of flux/force products, the previous equation finally allows us to obtain the driving forces associated with the plastic strains and the mater flux:

$$\begin{cases} F_{\dot{\varepsilon}_{ij}^{p}} = \sigma_{ij} \\ F_{\vec{J_{\gamma}}} = -\vec{\nabla} \left( \mu_{\gamma} + \sigma_{ij} \eta_{ij}^{\gamma} \right) + \left( \frac{\partial \sigma_{ij} \eta_{ij}^{\gamma}}{\partial c_{\gamma}} + \frac{\sigma_{ij} \eta_{ij}^{\gamma}}{c_{\gamma}} \right) \cdot \vec{\nabla} c_{\gamma} \end{cases} \tag{16}$$

From Eq. 16, we verify that in the case of stress-free diffusion ($\sigma=0$), the thermodynamic driving force associated with the diffusion is classically equal to the gradient of chemical potential. In what follows, we consider that the plastic strains are negligible, what is an acceptable hypothesis for a material showing an elastic behaviour, as the $UO_2$. Furthermore, noting that $\vec{\nabla} \mu_{\gamma} = \left( \partial \mu_{\gamma} / \partial c_{\gamma} \right) \vec{\nabla} c_{\gamma}$, and assuming the chemical potential definition given by Larché and Cahn in Eq. 17 [5], we obtain a final expression for the dissipation (Eq. 18) from which a flux law will ensue.

$$\mu_{\gamma} \left( \sigma_{ij}, T, c_{\gamma} \right) = \mu_{\gamma}^{c} \left( T, c_{\gamma} \right) - \sigma_{ij} \eta_{ij}^{\gamma}. \tag{17}$$

$$\mathcal{D}_{\delta V} = -\vec{J_{1}} \cdot \left\{ \left( \frac{\partial \mu_{1}^{c}}{\partial c_{1}} \right) - \left( \frac{\partial \sigma_{ij} \eta_{ij}^{1}}{\partial c_{1}} + \frac{\sigma_{ij} \eta_{ij}^{1}}{c_{1}} \right) \right\} \cdot \vec{\nabla} c_{1}. \tag{18}$$

In Eq. 17, $\mu_{\gamma}^{c} \left( T, c_{\gamma} \right)$ represents the chemical potential only depending on concentration and temperature. Oxygen diffusion in the system is higher than the one for U atoms, so we only consider the oxygen diffusion corresponding to eq. 18. The thermodynamic force driving the diffusion process is decomposed into a purely chemical part and a mechano-chemical part. From this equation, it is possible to obtain an evolution law looking like a Fick's law.

Let's now consider that:
- the fluxes are linear function of the forces (Onsager near–equilibrium conditions),
- the flux can be written in a classical way as $\vec{J_{1}} = -D_{1} \vec{\nabla} c_{1}$,
- the definition of the pure chemical potential is $\mu_{1}^{c} = \mu_{1}^{0} + RT \ln(x_{1})$, it comes:

$$D_{1}^{'} = D_{1} \left\{ 1 - \frac{\eta_{ij}^{1} c_{1}}{RT} \left( \frac{\partial \sigma_{ij}}{\partial c_{1}} + \frac{\sigma_{ij}}{c_{1}} \right) \right\}. \tag{19}$$

In the expression of the pure chemical potential, $x_1$ represents the molar fraction of oxygen in the system and $\mu_{1}^{0}$ its standard chemical potential. The diffusion coefficient $D_{1}^{'}$ can be separated into two components:
- The first one corresponds to the classical diffusion coefficient in a stress-free state: $D_{1}^{'} = D_{1}$. In this case, Eq. 19 corresponds to a classical Fick's law.
- The second one depends on both the stress state and the composition. This Nernst term conveys the forced diffusion induced by stresses during the diffusion of species inside the material. It directly influences the diffusion coefficient: according to the stress and/or its evolution, the oxygen dissolution will be speeded-up or slowed down.

**Numerical simulations**

$UO_2$ oxidation was chosen to perform numerical evaluation of our model because its characteristics make it a school case under many aspects.

First it is a simple system inducing a simplified numerical model
- as an oxide it has a pure elastic behaviour (no plastic term need to be taken into account).
- oxidation occurs by oxygen incorporation in the solid (the chemical state of the sample can be described by the oxygen concentration only).

Second it undergoes the chemical fragmentation phenomenon in several different configurations, which makes possible its phenomenological analysis. $UO_2$ oxidation generates indeed different crystalline phases as a function of temperature and oxidation rate:
- at temperature higher than 400°C, $U_3O_8$ phase is mainly formed with a 36% swelling compared to $UO_2$.
- At temperature lower than 300°C, a layer of $U_4O_9$ cubic phase is first formed with 0.5% shrinking compared to $UO_2$, then a layer of $U_3O_7$ tetragonal phase appears on $U_4O_9$ layer. $U_3O_8$ formation occurs afterwards.
- At temperature around 300°C, $U_4O_9$ formation is not observed and only a $U_3O_7$ layer exists at low oxidation rate, before $U_3O_8$ formation.

The values used for the different internal parameters are given in table 1.

|  | $D_1$ (cm³/s) | $\eta_{ij}^1/\rho$ | T [°C] | E [GPa] | ν | Lattice parameters [Å] | | |
|---|---|---|---|---|---|---|---|---|
|  |  |  |  |  |  | a | b | c |
| Ref. | [9] | [11] |  | [10] |  | [7,8] | | |
| $UO_2$ | $0.0055 \exp\left(\frac{-26.3}{RT}\right)$ | $-1.248 \cdot 10^{-5}$ | 300 | 200 | 0,32 | 5,47 (cubic) | | |
| $U_4O_9$ |  |  |  |  |  | 5,44 (cubic) | | |
| $U_3O_7$ |  |  |  |  |  | 5,40 (tetragonal) | | 5,49 |

Table 1: Parameters values used in the simulations (ρ corresponds to the density).

In this paper, we will focus on the early $U_3O_7$ formation at 300°C, which can be seen as an $U_3O_7$ layer in epitaxy on a $UO_2$ substrate. In this classical situation the mechanical state of the system, depends on the mismatch between the unit cell parameters of the substrate and the layer in the interface plane. Because $U_3O_7$ is a tetragonal phase, 2 different orientations of it unit cell are possible: either its c axis lye within the interface plane, or not. Because unit cell parameters verify $a(U_3O_7) < a(UO_2) < c(U_3O_7)$, $U_3O_7$ c axis parallel or perpendicular to the interface plane generates different stresses in $UO_2$ substrate.

The aim of the numerical calculations we performed was to identify what differences on the oxygen diffusion are induced by the different mechanical states created as a function of c axis orientation. For that purpose a simplified 2 dimensional geometry (Fig. 2) was used with the hypothesis of semi-infinite sample implying $c_1(h) = 0$ and $\sigma(h) = 0$ at the position h far from external interface (see Fig. 2). Oxygen penetrates the material through the face Σ and diffuses only on z direction. Because we focus on stress induced oxygen diffusion, the calculations were performed with a 1 s time scale, preventing the system to relax.

**Simulation of an oxide layer.** In the computation, the presence of an oxide layer is simulated, considering two simple configurations (presented in Fig. 2):
- in the case (B), the oxide lattice parameter parallel to the sample surface is smaller than the substrate's one. To link the two lattices, it is necessary to expand the oxide layer. This expansion leads to compressive $\sigma_{xx}$ stress on the substrate surface.
- in the case (C), the oxide lattice parameter parallel to the sample surface is higher than the substrate's one. The linking of the two lattices induces a tensile $\sigma_{xx}$ stress on the substrate surface.

As comparison, case (A) gives the material response without oxide layer, i.e. without induced stress.

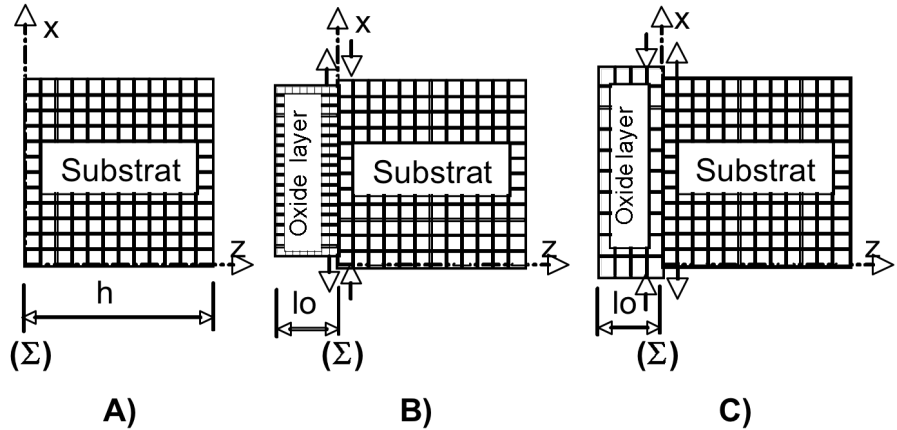

Fig. 2: The different stress fields applied to the sample at internal interface ($l_0=1\mu m$).

**Theoretical results.** Fig. 3 gives, as a function of the distance to the interface $\Sigma$:
- the evolution of oxygen concentration in the material,
- the evolution of the internal stress.

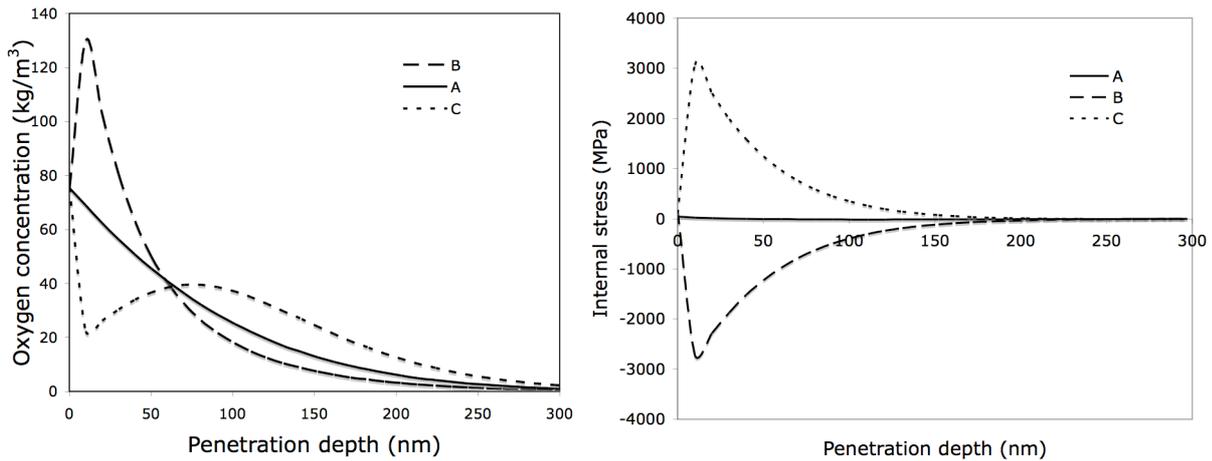

Fig. 3: Oxygen concentration and internal stress evolutions vs. penetration depth into the substrate for three different imposed strain fields.

Even if it is still difficult to confirm experimentally the reached level of stresses calculated in these simulations, because of interface $\Sigma$ does not move in this approach, it is interesting to observe strong behaviour differences between the two simulated cases. If the formed oxide layer tends to distend the substrate's crystalline lattice (positive tensile stress, what corresponds to case C), we notice that oxygen concentration slumps near the interface $\Sigma$ and increases when penetrating $UO_2$. This shows a slow down of the oxidation process due to strong oxygen dissolution in the volume. On the contrary, if the formed oxide layer compresses the crystalline lattice (negative stress, which corresponds to case B), the oxygen cumulates near the interface $\Sigma$.
In the case B, the oxide layer formation is made easier than in the case C, because the oxygen saturation is rapidly reached near the surface.

In a 3 dimensional configuration, experimental observations on uranium dioxide oxidation [6] show that the $U_3O_7$ phase has its (011) plane parallel to the surface. This situation corresponds to a case more complex than case B and C. $UO_2$ with a 5.47*5.47 Å² cell surface has to fit $U_3O_7$ with a 5.40*5.49 Å² cell surface. Taking only cell surface into account it would correspond to case B. In fact $U_3O_7$ is created with domain formation and the orientation of c axis is not identical in the domains, which justify the hypothesis taking into account only cell surface in order to describe $UO_2/U_3O_7$ interface.

As a consequence, the oxygen diffusion profile calculated in case B is consistent with the assumption made in literature [12], according which oxidation proceeds thanks to the progression of the oxidised phase into $UO_2$ with a sharp interface and nearly no oxygen diffusion in $UO_2$. Concerning the case C, the very original behaviour it induces in $UO_2$ deserves to be analysed more precisely, in comparison with the behaviours observed experimentally in the formation of the different phases inside the oxide layer.

**Conclusion**

A model is proposed to explain firstly the appearance of stresses at the substrate/oxide interface during the material oxidation, and secondly the influence of these stresses on the oxidation process evolution. For this occasion, a new expression of the diffusion coefficient, taking into account the stresses, is done.

This work is a study of sensitivity, in which it was shown that the evolution of a system substrate/oxide largely depends of the strains induced by the interfacial chemical transformation. Through the equations applied to the $UO_2$ oxidation study, the calculated profiles of concentration (case B) can be interpreted in the literature. We observe also, in the simulations, very different behaviours on whether the stress induced in the substrate is negative or positive.

This first mechano-chemical approach of the diffusion mechanisms into a solid subjected to a strain field will be followed by a more detailed reconstruction of the material oxidation kinetic. This second step of our approach will be carried out soon, and based on a thermodynamic study of the solid/solid reactive interface displacement speed. This dynamic study will be completed by a static study of a multi-layer system ($UO_2/U_4O_9/U_3O_7$), which will allow us to evaluate the crystallographic compatibility of the lattices with regards to the reactional schemes proposed in the litterature. For this occasion, the strains generated by the linking of the different lattices will be studied thanks to the Bollmann's method.